\documentclass[runningheads]{llncs}
\usepackage[T1]{fontenc}
\usepackage{graphicx}
 \usepackage{dsfont}
\begin{document}
\title{SuperFormer: Volumetric Transformer Architectures for MRI Super-Resolution 
}
\titlerunning{SuperFormer}
%
\author{Cristhian Forigua \inst{1}\orcidID{0000-0003-4472-1144} \and Maria Escobar \inst{1}\orcidID{0000-0002-5880-762X} \and Pablo Arbelaez \inst{1}\orcidID{0000-0001-5244-2407}}
\authorrunning{C. Forigua et al.}
%
\institute{Center for Research and Formation in Artificial Intelligence, Universidad de los Andes, Bogotá, Colombia\\
\email{\{cd.forigua, mc.escobar11, pa.arbelaez\}@uniandes.edu.co}}
\maketitle              
\begin{abstract}
This paper presents a novel framework for processing volumetric medical information using Visual Transformers (ViTs). First, We extend the state-of-the-art Swin Transformer model to the 3D medical domain. Second, we propose a new approach for processing volumetric information and encoding position in ViTs for 3D applications. We instantiate the proposed framework and present SuperFormer, a volumetric transformer-based approach for Magnetic Resonance Imaging (MRI) Super-Resolution. Our method leverages the 3D information of the MRI domain and uses a local self-attention mechanism with a 3D relative positional encoding to recover anatomical details. In addition, our approach takes advantage of multi-domain information from volume and feature domains and fuses them to reconstruct the High-Resolution MRI. We perform an extensive validation on the Human Connectome Project dataset and demonstrate the superiority of volumetric transformers over 3D CNN-based methods. Our code and pretrained models are available at \url{https://github.com/BCV-Uniandes/SuperFormer} 
\end{abstract}
\section{Introduction}
High Resolution (HR) Magnetic Resonance Imaging (MRI) contains detailed anatomical structures that are crucial for accurate analysis and diagnosis of several diseases. This modality is commonly used in specialized medical centers. Still, its broader adoption is hindered by costly equipment and long scan times, resulting in small spatial coverage and low Signal-to-Noise Ratio (SNR) \cite{degradacion}. In contrast, Low Resolution (LR) imaging requires lower acquisition time, less storage space, and less sophisticated scanners \cite{lowcost}. Nevertheless, LR medical images suffer from artifacts of patient's motion and lack detailed anatomical structures. Therefore, volumetric image Super-Resolution (SR) is a promising framework for generating HR MRIs by mapping them from a LR input while keeping the advantages of the LR acquisition.
\begin{figure}[t]
    \centering
    \includegraphics[width=0.9\textwidth]{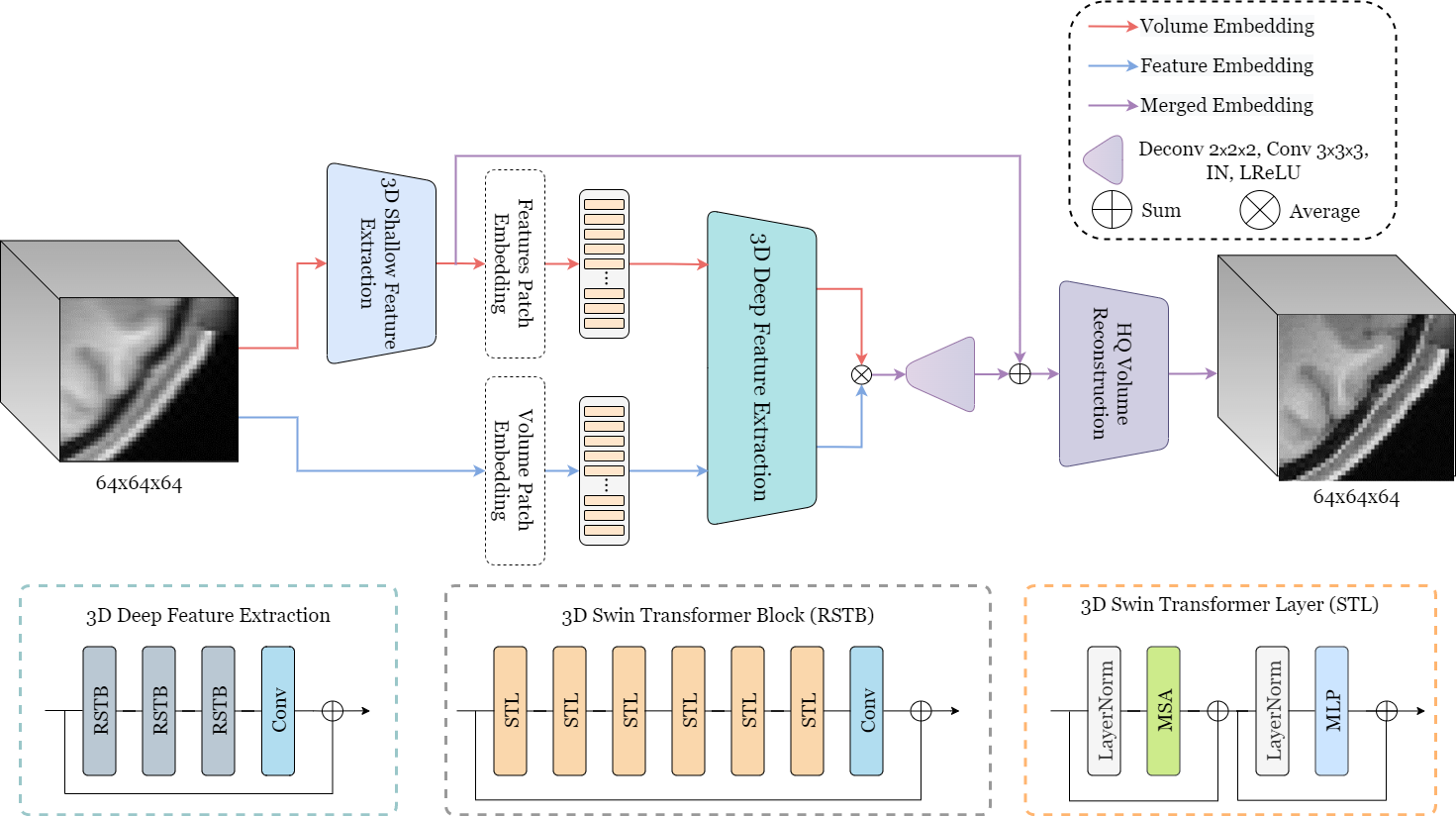}
    \caption{Overview of our method. SuperFormer encodes features and volume embeddings for deep feature extraction through volumetric transformers and combines the multi-domain representations to reconstruct the super-resolved volume.}
    \label{overall}
\end{figure}

The human Connectome Project (HCP) dataset \cite{HCP} has been widely used to study the medical imaging SR framework since it provides HR volumetric MRI from 1,113 subjects \cite{miccai2018,degradacion,wang}. There are two different approaches that tackle the problem of medical imaging SR. On one hand, 2D oriented methods interpret the 3D information as a set of slices without any volumetric interaction \cite{MRI2d_2,MRI2d_1}. On the other hand, 3D-based approaches exploit the inherent connections of the volumetric information \cite{MRI3d_1,MRI3d_2,MRI3d_3,MRI3d_5}. For instance, the seminal works mDCSRN \cite{miccai2018} and MRDG48 \cite{wang} propose 3D Generative Adversarial Networks (GANs) to super-resolve MRI volumes with high perceptual quality on the HCP dataset.   However, there is a lack of a standardized evaluation framework for 3D medical image SR, and the proposed methods are not publicly available, limiting the progress in this area.

 Recently, the computer vision community has experienced the revolution of Vision Transformers (ViT)s \cite{vit} for several computer vision tasks \cite{DETR,mask2former,swinTransformer}. Unlike CNNs, which have restricted receptive fields, ViTs encode visual representations from a sequence of 2D patches and leverage self-attention mechanisms for capturing long-range global information \cite{UNETR}. Since ViTs learn stronger feature representations, they are commonly used as feature encoders for further downstream tasks. Liu \textit{et al.} proposed Shifted windows (Swin) Transformer, a 2D transformer backbone that uses non-overlapping windows to compute self-attention at multi-scale modeling. Since its introduction, Swin Transformers have achieved state-of-the-art results in several computer vision tasks \cite{swinTransformer}. Although transformers are being adopted in the medical imaging domain for problems such as 3D medical segmentation \cite{UNETR,karimi2021convolutionfree,swinUNETR,Seg-Trans,zhang2021multibranch}, their potential remains unexplored on the 3D medical SR framework. Feng \textit{et al.} \cite{TaskTransformer} proposed Task Transformer Network, a transformer-based method for MRI reconstruction and SR. However, this 2D-oriented approach does not fully leverage the continuous information in the 3D domain since they process volumes on a slice by slice basis, ignoring the inherent volumetric information of MRIs.  

In this paper, we present SuperFormer, a volumetric approach for MRI Super-Resolution based on 3D visual transformers. Inspired by the success of ViTs for image restoration tasks \cite{swinir}, we create a volumetric adaptation for medical imaging. This approach is the first to use ViTs in the 3D domain for medical imaging Super-Resolution to the best of our knowledge. Figure \ref{overall} shows an overview of our method. We leverage the 3D and the multi-domain information from the volume and feature embeddings. Moreover, our approach uses a local self-attention mechanism with a 3D relative position encoding to generate SR volumes by volumetric processing. We perform an extensive validation and demonstrate the superiority of volumetric transformers over 3D CNN-based methods. Additionally, our results indicate that using multi-domain embeddings improves the performance of volumetric transformers compared to single-domain approaches. Our main contributions can be summarized as follows: 
\begin{enumerate}
    \item We propose a 3D generalization of the Swin Transformer framework to serve as a general-purpose backbone for medical tasks on volumetric data.
    \item We introduce a new approach for processing volumetric information and encoding position in ViTs for 3D frameworks, increasing the transformer's receptive field and ensuring a volumetric understanding. 
    \item We present SuperFormer, a novel volumetric visual transformer for MRI Super-Resolution. Our method leverages the 3D and multi-domain information from volume and feature embeddings to reconstruct HR MRIs using a local self-attention mechanism. 

\end{enumerate}
Furthermore, we provide a medical SR toolbox to promote further research in 3D MRI Super-Resolution. Our toolbox includes the extension of one transformer-based and two CNN methods into the 3D framework. Our toolbox, source code, dataset division, and pre-trained models are publicly available.  
\section{Method}
We propose SuperFormer, a transformer-based network that generates HR MRI volumes by mapping them from a LR volume to the high-resolution space. SuperFormer leverages the 3D information of medical images and multi-domain information by using volumetric transformers that analyze the image and feature domains of the input volume. Then, SuperFormer processes both domains together to reconstruct the HR output. Fig. \ref{overall} shows an overview of the proposed method.

\subsubsection*{Volumetric Transformer Processing}
Transformer architectures for natural images operate on a sequence of 1D input embeddings extracted from the 2D domain. In the 3D domain, we need to compute these embeddings from volumetric information. Given an input volume $X \in \mathds{R}^{H\times W\times D\times C}$, we  first extract 3D tokens as non-overlapping patches with a resolution of $(H',W', D')$ from $X$. Each token has a dimension of $H'\times W'\times D' \times C$. Subsequently, we project the 3D tokens into a $C_{emb}$-dimensional space, which remains constant throughout the transformer layers via an embedding layer. Then, we flatten the embedded representations to obtain the one-dimensional embeddings that the transformers process. Under this configuration, the transformers can operate on 1D embeddings that encode the 3D information and provide a volumetric perception of the input volume, have a larger receptive field depending on the patch resolution, and have fewer computational costs and hardware constraints. 
\subsection{Feature Embedding}
To compute the feature representation of the input volume, SuperFormer first extracts 3D shallow features directly from the input. Then, the feature patch embedding layer extracts 3D tokens from these features and projects them for further volumetric transformer processing. Since the shallow features encode low-frequency information of the input volume, the Feature Embedding ensures the transformers consider the low frequencies for recovering the lost anatomical details.  
\subsubsection*{3D Shallow Feature Extraction}
Given a LR volume $V_{LQ} \in \mathds{R}^{H\times W \times D \times C_{in}}$, where $H$, $W$, $D$ and $C_{in}$ are the image height, width, depth and input channel number, respectively, we use a $3\times3\times3$ convolutional layer to extract the shallow features $F_0 \in \mathds{R}^{H\times W \times D \times C_{emb}}$, where $C_{emb}$ is the embedding dimension. By putting this layer at an early stage of processing, we achieve better results and more stable optimization \cite{early_conv}, while mapping the input channel dimension to a higher dimensional space for volumetric processing.
\subsection{Volume Embedding}
The intensity information in the 3D domain contains relevant information about patterns and anatomical structures that are volumetrically organized in the MRI. Thus, encoding the volume domain provides a clear idea of the structures we want to recover. SuperFormer encodes volume representations by processing the input directly in the volume domain. We use a volumetric patch embedding layer that computes the 3D tokens directly from the input and projects them into the $C_{emb}$-dimensional space.

\subsection{3D Deep Feature Extraction} SuperFormer processes the feature and volume embedding representations by extracting 3D deep features for each domain. To compute these deep representations, SuperFormer employs $K$ 3D Residual Swin Transformer blocks (RSTB) and a final $3 \times 3 \times 3$ convolutional layer. Each RSTB is a residual block that consists of $L$ 3D Swin transformer layers (STL), a $3 \times 3 \times 3$ convolution layer, and a residual connection. 
\subsubsection*{3D Swin Transformer layer.} 
\begin{figure}[t]
    \centering
    \includegraphics[scale=0.3]{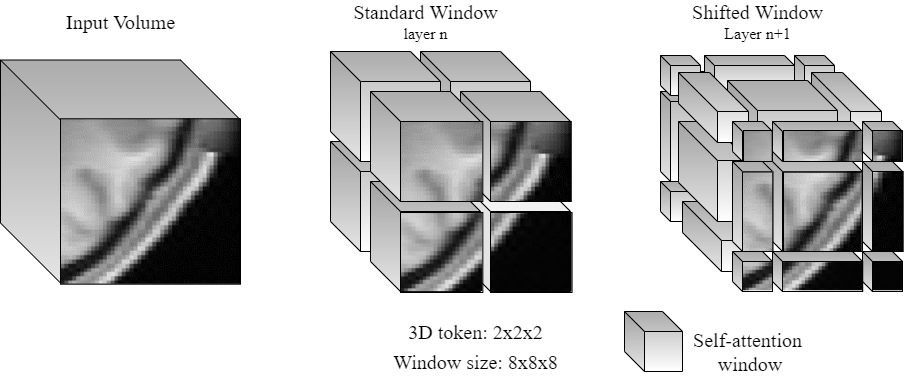}
\caption{3D Shifted window for computing self-attention in the Deep Feature Extraction for 2$\times$2$\times$2 3D token and 8$\times$8$\times$8 window size.}
    \label{shifted mechanisim}
\end{figure}

In contrast to the standard multi-head self-attention of the original visual transformer \cite{mvit}, we model volumetric interactions between tokens as three-dimensional local self-attention with a 3D shifted window mechanism \cite{swinTransformer}. Then, we partition the input volume into non-overlapping windows to compute the local self-attention, as shown in Fig. \ref{shifted mechanisim}. Precisely, at STL $n$, we compute windows of size $M \times M \times M$ to  split the input 3D token into $[\frac{H'}{M}] \times [\frac{W'}{M}] \times [\frac{D'}{M}]$ regions. Then, at STL $n+1$, we shift the previous windows by $\frac{M}{2}$ voxels along each dimension. Inspired by \cite{swinTransformer,swinUNETR}, we adopt a 3D cyclic-shifting for the continuous computation of the shifted partitions. 
Moreover, for a feature in a local window we compute the \textit{query (Q)}, \textit{key (K)} and \textit{value (V)} matrices as
\begin{equation}
    Q=XP_Q,\hspace{1cm} K=XP_K,\hspace{1cm} V=XP_V
\end{equation}
where $X$ is the local window feature, $P_Q, P_K$ and $P_V$ are the shared projection matrices between all windows. We then compute the self-attention as
\begin{equation}
    Attention(Q,K,V) = Softmax(\frac{QK^T}{\sqrt{d+B}})V.
\end{equation}
where $B$ is a relative position encoding for learning the locations of the 3D tokens regarding the input volume. Unlike most ViTs-based models for 3D image processing \cite{UNETR,swinUNETR}, which use absolute position embeddings \cite{mvit}, we adapt the 2D absolute position bias to the 3D framework. This positional encoding has been proved to lead to better results in the natural image domain \cite{swinir,swinTransformer}.
The whole STL configuration, as shown in Fig. \ref{overall}, consists of the standard or shifted local attention mechanism, normalization layers, and a Multi-Layer Perceptron (MLP).
\subsection{HQ Volume Reconstruction.}
SuperFormer learns a deep representation from the shallow features and volume domains to reconstruct the SR volume. First, we combine the information of the two domains by averaging the deep features and aggregating the shallow features with a residual connection. Since deep features differ in size because of the volumetric processing, we employ an up-sampling layer to match the dimensions. Intuitively, the shallow features convey low-frequency information, while the deep features concentrate on recovering the high frequencies. The residual connection transfers the low frequencies directly into the reconstruction stage to recover high frequencies. Specifically, we reconstruct the HR volume by a $3\times3\times3$ and $1\times1\times1$ convolutional layers with a LeakyReLU activation function. 
\section{Experimental setup}
\textbf{Dataset} We use the Human Connectome Project (HCP) dataset \cite{HCP} to train and test our method, a large publicly accessible brain structural MRI dataset containing high spatial resolution 3D T1W images from 1,113 healthy subjects. All the images were acquired via Siemens 3T platform using a 32-channel head coil on multiple centers and come in high spatial resolution as 0.7 mm isotropic with a matrix size of 320x320x256. Ground truth annotations come from these high-quality images, which provide detailed anatomical structures. We used the same data distribution as in \cite{miccai2018,wang}; 780 for training, 111 for validation, 111 for evaluation, and 111 for testing. We calculate three evaluation metrics for quantitatively measuring the similarity between HR and SR volumes: Peak Signal to Noise Ratio (PSNR), Normalized Root Mean Squared Error (NRMSE), and subject-wise average Structural Similarity Index (SSIM).\\
\textbf{Low-Resolution Volumes Generation} Since we need a paired dataset of LR and HR images to train and test SR models, we generate the LR images from the HR ones by following the approach used in \cite{degradacion}. First, we convert the HR volumes into k-space by applying the Fast Fourier Transform (FFT). Then, we degrade the resolution by truncating the outer part of the 3D k-space by a factor of 2x2. Finally, we convert back to the image domain by applying the inverse FFT and linear interpolation to the original HR volume size. This approach mimics the actual acquisition process of LR and HR MRIs since it does not perform any dimensional change but undersamples the frequency k-space \cite{miccai2018}.
\subsection{Implementation Details}
We implement SuperFormer in PyTorch and train the model on a workstation with 4 Nvidia RTX 8000 GPUs during 55k iterations during four days. We use an ADAM optimizer with a learning rate of $2\mathrm{e}{-4}$ to minimize the L1 loss function with a batch size of 4. For the ablation studies, we use 252 as embedding dimension, six RSTBs, each with six STLs and six attention heads to compute the local self-attention. In SuperFormer's configuration, we use three RSTBs instead of six because of hardware constraints. In addition, we use a patch resolution of $2\times2\times2$ for volumetric processing.
\subsection{Results}
\subsubsection*{Comparison with the State-of-the-art} 
We evaluate SuperFormer against two CNN-based methods: EDSR  \cite{cnn6}, and RRDBNet \cite{gan2}. Since these methods are originally designed for 2D SR, we extend them for the 3D domain. In addition, we compare our method against the 2D transformer-based SwinIR \cite{swinir} to demonstrate the advantages of leveraging volumetric information. We can not directly compare our approach against mDCSRN \cite{miccai2018} nor MRDG48 \cite{wang} because there is no publicly available code for these methods and their experimental framework is different.

Table \ref{comparison} shows the results in our test split of the HCP dataset. First, by comparing SuperFormer and SwinIR against 3D EDSR and 3D RRDBNet, we find that using a transformer-based approach is greatly beneficial for the task of SR. In fact, SuperFormer outperforms both 3D CNN methods in the PSNR and NRMSE metrics. Moreover, using a 2D transformer like SwinIR still produces higher results than the 3D CNN counterparts. These findings validate the superiority of transformers for deep feature extraction in SR. 
Second, the comparison between SuperFormer and SwinIR empirically demonstrates that leveraging 3D information through volumetric transformers improves the performance of MRI SR. This finding is consistent with our intuition since the volumetric transformer interprets a three-dimensional vision of the anatomical structures and acquires a broader context for generating a more coherent SR output. Furthermore, Superformer performs significantly better (p-value<0.05) and has fewer parameters.

\begin{figure}[t]
    \centering
    \includegraphics[width=1.0\textwidth]{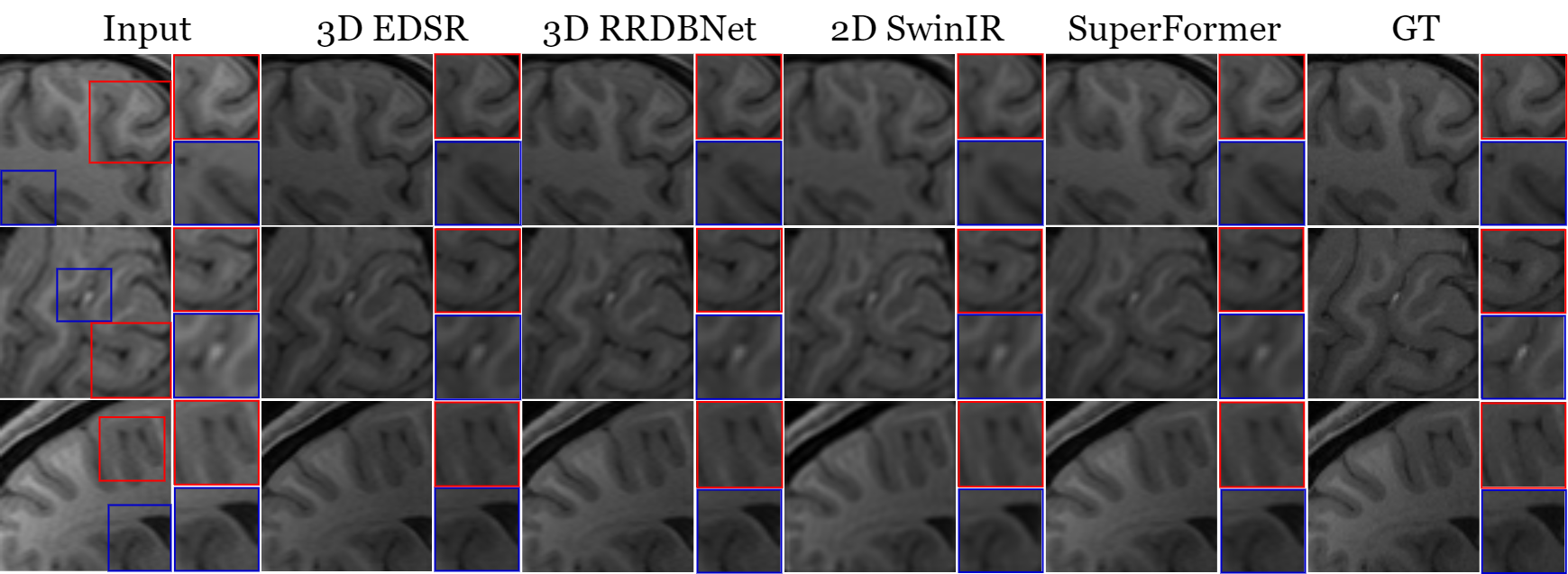}
    \caption{Qualitative comparison of our method against CNN and 2D transformer-based methods on the axial, coronal and sagittal anatomical axes.}
    \label{qualitative}
\end{figure}

Fig \ref{qualitative} shows qualitative results that support the findings of Table \ref{comparison}. In particular, the improvement in perceptual quality is noticeable when comparing the 2D SwinIR baseline against our implementation of volumetric transformers. The images super-resolved with volumetric transformers have sharper structures in comparison with their 2D SwinIR counterpart.
\begin{table}[t]
\centering
\caption{Comparison of our method against the state-of-the-art methods on the test split. SuperFormer outperforms 3D CNN and transformer-based approaches.}
\label{comparison}
\begin{tabular}{c|c|c|c|c}
                                         & \textbf{\#Params} & \textbf{PSNR $\uparrow$} & \textbf{SSIM $\uparrow$} & \textbf{NRMSE $\downarrow$} \\ \hline
2D SwinIR \cite{swinir} & 22.5M
 & 31,4285$\pm$3,5496        & 0,8297$\pm$0,0303         & 0,2002$\pm$0,0761          \\
3D EDSR & 41.2M                             &31,0217$\pm$3,1195                    & 0,9147$\pm$0,0314                    &0,2050$\pm$0,0636                       \\
3D RRDBNet & 115M                           & 31,3059$\pm$3,3876                  & \textbf{0,9355$\pm$0,0207}                    & 0,2006$\pm$0,0692                       \\ \hline
\textbf{SuperFormer} & \textbf{20M}
 &
  \textbf{32,4742$\pm$2,9847} &
  0,9059$\pm$0,0271&
  \textbf{0,1747$\pm$0,0635}
\end{tabular}
\end{table}
\subsubsection*{Ablation Study} 
Our ablation studies include a thorough analysis of SuperFormer's embedding configuration. We compare our method against three different configurations: \textit{SR-Features} when we compute the deep features only from the shallow features embedding, \textit{SR-Volume} when we extract them just from the volume embedding, and \textit{SR-Avg} when we take the average between the volume and feature embeddings to compute the deep representations.

Table \ref{ablation} shows the results of our final model and the ablation experiments for the validation set of the HCP dataset. These results empirically validate the relevance of using multi-domain embeddings in our final method. If we use the embedding representation of only one domain, regardless of which one, the PSNR decreases by 1 decibel. Additionally, we explore how our approach merges the two embedding representations. The results demonstrate that averaging the extracted deep features brings an improvement across all metrics compared to averaging the embedding representations. These ablation experiments show that SuperFormer's embedding configuration is beneficial for 3D MRI SR. 
\begin{table}[]
\centering
\caption{Ablation experiments for SuperFormer's branch configuration on the HCP validation set. We report the results of our final method and ablation experiments.}
\label{ablation}
\begin{tabular}{c|c|c|c}
                                    & \textbf{PSNR $\uparrow$} & \textbf{SSIM $\uparrow$} & \textbf{NRMSE $\downarrow$} \\ \hline
\textit{SR-Volume} & 31,5354$\pm$2,145            & 0,8644$\pm$0,015             & 0,1829$\pm$0,035\\
\textit{SR-Features}          & 31,5195$\pm$3,5209                           & \textbf{0,9085$\pm$0,344} & 0,1906$\pm$0,0659                         \\
\textit{SR-Avg}    & 32,0557$\pm$3,1081             & 0,8950$\pm$0,0294              & 0,1777$\pm$0,0570                \\ \hline
\textbf{SuperFormer} &
  \textbf{32,5164$\pm$3.028} &
  0,9059$\pm$0.027&
  \textbf{0,1668$\pm$0.046}
\end{tabular}
\end{table}
\section{Conclusion}
In this work, we present SuperFormer, a method for SR that leverages the 3D information of MRIs by using volumetric transformers. To the best of our knowledge, SuperFormer is the first method for this task that implements 3D visual transformers and uses a volumetric local self-attention mechanism. We experimentally demonstrate that leveraging the 3D information through volumetric transformers and multi-domain embeddings leads to better results compared to state-of-the-art approaches. Furthermore, our generalizable framework will push the envelope further in the development of volumetric Visual transformers for medical imaging.


%
%
%
%
\bibliographystyle{splncs04}
\bibliography{bibliography}
\end{document}